\documentstyle[preprint,aps]{revtex}

\newcommand{\be}{\begin{equation}}

\newcommand{\ee}{\end{equation}}

\newcommand{\bea}{\begin{eqnarray}}

\newcommand{\eea}{\end{eqnarray}}

\newcommand{\e}{\mbox{{\rm e}}}                     

\newcommand{\Eq}{Eq.}

\newcommand{\eqs}{Eqs.}
   
\newcommand{\sgn}{{\rm sgn}}

\newcommand{\Li}{\mbox{\rm Li}}

\newcommand{\ts}{\tilde{\sigma}}

\newcommand{\tr}{\tilde{\rho}}

\begin{document} 

\draft

\title{\large\bf Symmetry Nonrestoration in a Gross-Neveu Model  
                \\ with Random Chemical Potential}

\author{Seok-In Hong$^{1,2}$ and John B. Kogut$^1$}

\address{$^1$Loomis Laboratory of Physics,
          University of Illinois at Urbana-Champaign\\
          1110 West Green Street, Urbana, IL 61801-3080\\   
           $^2$Department of Science Education,
           Inchon National University of Education\\ 
           Inchon, 407-753, Korea}

\date{\today}

\maketitle

\begin{abstract}

We study the symmetry behavior of the Gross-Neveu model 
in three and two dimensions with random chemical potential. 
This is equivalent to a four-fermion
model with charge conjugation symmetry as well as $Z_2$ chiral
symmetry. At high temperature the $Z_2$ chiral symmetry is always restored.
In three dimensions the initially broken charge conjugation symmetry is not
restored at high temperature, 
irrespective of the value of the disorder strength. 
In two dimensions and at zero temperature the
charge conjugation symmetry undergoes a quantum phase transition from
a symmetric state (for weak disorder) to a broken state (for strong disorder) 
as the disorder strength is varied. For any given
value of disorder strength, the high-temperature behavior of the charge 
conjugation symmetry is the same as its zero-temperature behavior.
Therefore, in two dimensions and for strong disorder strength 
the charge conjugation symmetry is not restored at high temperature.
\end{abstract}


\narrowtext

\section{Introduction}

Intuitively, when heated, a system with initially broken symmetry will 
recover its symmetry because thermal fluctuations are able to overcome
potential barriers. But a counterexample was noticed by Weinberg
\cite{weinberg}: For the four-dimensional O($N$)$\times$O($N$) scalar $\phi^4$
model, he showed that the system can remain in the broken phase 
even at sufficiently high temperature. This phenomenon is called inverse
symmetry breaking or symmetry nonrestoration (SNR), depending on whether the 
system was in a symmetric or a broken phase at zero temperature.

Since Weinberg's observation, SNR has
been a subject of academic curiosity or a candidate way out of cosmological
problems caused by topological defects like monopoles 
and domain walls (see Ref.\cite{bajc} for review). 
According to Bajc's classification\cite{bajc}, there are three classes
of SNR mechanisms in field theory: (i) a prototype case like the two-scalar
model\cite{weinberg,prototype}, (ii) flat directions in supersymmetric 
theories\cite{flat}, and
(iii) large charge density (or chemical potential)\cite{chemical,charge}. 
Here we restrict ourselves to class (iii).

If a large enough charge can not be stored in thermally excited modes at high
temperature, it must reside in the vacuum, and this is a sign of SNR.
In field theory, a scalar field (order-parameter field) gets a positive mass
term by thermal effects, but a negative one by the effects due to the 
chemical potential. For a fixed charge (i.e., in the canonical formalism),  
the chemical potential is temperature dependent. 
In this case, if the effect of the chemical potential on the mass
exceeds the thermal effects at sufficiently high temperature, the scalar field 
acquires a nonzero vacuum expectation value (i.e., SNR)\cite{chemical}.

However, in an open system of the model which does not belong to
the class (i) or (ii), the symmetry may always be restored at
high temperature. In the grand canonical formalism, the chemical potential 
and the temperature are independent parameters and so the thermal effect
on the mass always surpasses the effect of the chemical potential 
at sufficiently high temperature, for a fixed chemical potential. 
For example, consider the Gross-Neveu (GN) model
\cite{gn,rwprep,hkkann} with chiral symmetry. At finite chemical potential
the initially broken chiral symmetry is always restored 
at high temperature\cite{rwphkk}.

In order to find a new kind of SNR in four-fermion models, 
we will extend the GN model at finite chemical potential 
to a disordered model with random chemical potential. Recently,
disordered nonrelativistic Dirac fermions in two spatial dimensions have 
been studied in relation to the integer quantum Hall transition\cite{iqht}. 
Pure fermions exhibit such a transition 
as the value of the mass is tuned through
zero, but its universality class is different from the one observed in actual
experiments. Usually three types of (static) disorder are considered for a more
realistic model: random gauge potential, random chemical potential and random
mass.

Motivated by the SNR mechanism (iii), we introduce the
(relativistic) GN model with random chemical potential in Sec. II.
If the chemical potential has a Gaussian distribution at each
site, our model is equivalent to the four-fermion model with two
kinds of four-fermion interaction, $(\bar{\Psi}\Psi)^2$ and $(\bar{\Psi} 
\gamma_0 \Psi)^2$ (see \Eq (3)), and has charge conjugation symmetry in 
addition to the $Z_2$ chiral symmetry. In Sec. III we examine 
the behavior of these symmetries as the temperature or disorder strength is 
varied using the $1/N$ expansion in three and two dimensions.
While $Z_2$ chiral symmetry is always restored at high temperature, the
charge conjugation symmetry exhibits SNR. In addition, we check the validity
of the mean field approximation (the leading approximation in the $1/N$
expansion) in two dimensions. In Sec. IV the fundamental origin
of SNR for charge conjugation symmetry is discussed conceptually.
Our conclusions are presented in Sec. V.

\section{Gross-Neveu model with random chemical potential} 

The Euclidean Lagrangian of the GN model at finite chemical potential $\mu$ is
given by
\be
{\cal L} = \bar{\Psi}(\not\partial+\mu \gamma_0 ) \Psi-\frac{g^2}{2N} 
(\bar{\Psi} \Psi)^2 ,
\ee
where $g^2(>0)$ is the coupling constant of the four-fermion interaction 
$(\bar{\Psi}\Psi)^2$ and $N$ 
is the number of flavors of the Dirac fermion $\Psi$. The $\gamma$ matrices are
$4 \times 4$ and hermitian. Let us consider the system under the influence of a random 
chemical potential $\rho (x)$ with the Gaussian distribution
$\exp(-\int d^d x \frac{N}{2 R^2} \rho^2) $ 
at each site where $R^2 (>0)$ is the strength
of disorder and $d$ the dimension of the Euclidean space. The Gaussian
noise\cite{zinnjustin} is characterized by correlation functions
\be
\langle \rho (x) \rangle = 0,~~
\langle \rho (x) \rho (x^\prime )\rangle = \frac{R^2}{N} \delta^d 
(x-x^\prime ).
\ee 

After integrating out the random chemical potential, our model
is equivalent to the four-fermion model
\be
{\cal L}=\bar{\Psi} \not\partial \Psi-\frac{1}{2N} \left[ g^2 (\bar{\Psi}
\Psi )^2 +R^2 (\bar{\Psi} \gamma_0 \Psi)^2 \right],  
\ee
with the $Z_2$ chiral symmetry $\{ \Psi \rightarrow \gamma_5 \Psi,~ 
\bar{ \Psi }\rightarrow - \bar{\Psi} \gamma_5 \}$ 
and the charge conjugation symmetry
$\{ \Psi \rightarrow C \bar{\Psi}^T ,~ \bar{\Psi} 
\rightarrow - \Psi^T C^\dagger\}$.
Here the matrix $C$ satisfies $C^\dagger C =1,~C^\dagger \gamma_\mu C
= - \gamma_\mu^T$. Under charge conjugation, $\bar{\Psi} \Psi$ and 
$\bar{\Psi} \gamma_0 \Psi$ transform to $\bar{\Psi} \Psi$ and 
$-\bar{\Psi} \gamma_0 \Psi$ respectively. Hence, the Lagrangian \Eq (1) with 
definite chemical potential $\mu$ does not possess the charge conjugation 
symmetry (i.e., fermion-antifermion symmetry).
Note that in \Eq (3) the chemical potential term does not appear explicitly.  

We will study this model by the leading approximation of the $1/N$ expansion 
in three and two dimensions. To 
easily incorporate the $1/N$ expansion, let us rewrite the 
Lagrangian \Eq (3) by introducing scalar auxiliary fields $\sigma (x)$ and
$\rho (x)$:
\be
{\cal L}=\bar{\Psi} (\not\partial +\sigma+\rho \gamma_0 ) \Psi  +\frac{N}{2 g^2} \sigma^2 +\frac{N}{2 R^2}
\rho^2.
\ee
The random chemical potential $\rho (x)$ plays the role of a scalar auxiliary
field. The $Z_2$ chiral symmetry and the charge conjugation symmetry are now
expressed as $\{ \Psi \rightarrow \gamma_5 \Psi , \bar{\Psi} \rightarrow
-\bar{\Psi} \gamma_5 , \sigma \rightarrow -\sigma \}$ and
$\{ \Psi \rightarrow C \bar{\Psi}^T , \bar{\Psi} \rightarrow
-\Psi^T C^\dagger , \rho \rightarrow -\rho \}$, respectively.

\section{The behavior of $Z_2$ chiral symmetry and charge conjugation 
symmetry at zero and high temperature}

For finite-temperature field theory we adopt the imaginary-time formalism. 
At inverse temperature $\beta(= T^{-1})$, the fermion fields are
antiperiodic on $R^{d-1} \times [0,\beta ]$, while the scalar auxiliary fields
are periodic. Let us introduce the notation: $\int^{(T)}_{p} \equiv T 
\sum^{\infty}_{n=-\infty} \int \frac{d^{d-1} {\bf p}}{(2 \pi )^{d-1}}, ~ 
\int^{(0)}_{p} \equiv \int \frac{d^d p}{(2 \pi)^d }$. Integrating out the
fermion fields in the partition function for \Eq (4) 
we obtain the effective action for the auxiliary
fields $\sigma$ and $\rho$. In order to investigate the vacuum structure we
need to find the finite-temperature effective potential 
$V_T (\sigma , \rho )$ by taking $\sigma$
and $\rho$ as constant fields: To leading order in the $1/N$ expansion,
\be
\frac{V_T (\sigma, \rho)}{N} = \frac{\sigma^2}{2 g^2}+\frac{\rho^2}{2 R^2}
-2  \int^{(T)}_p \ln [ (p_0 - i \rho)^2 +{\bf p}^2 +\sigma^2 ],
\ee
where $p_0 = (2n+1)\pi /\beta \equiv \omega_n~(n={\rm integer})$ at nonzero
temperature. Note that the effect of the chemical potential $\rho$ is to shift
the energy by $-i\rho$.

To evaluate the integration in \Eq (5), we need some mathematical formulae.
By the standard method of contour integration\cite{thermal},
\bea
T \sum^{\infty}_{n=-\infty} \frac{1}{(\omega_n -i\rho)^2+\sigma^2 }
&=& \frac{1}{2 |\sigma|} \left[ 1-\frac{1}{1+\e^{\beta(|\sigma|+|\rho|)}}
-\frac{1}{1+\e^{\beta(|\sigma|-|\rho|)}} \right], \\
T \sum^{\infty}_{n=-\infty} \frac{\omega_n -i\rho }{(\omega_n -i\rho)^2
+\sigma^2 }&=& i\rho T \sum^{\infty}_{n=-\infty} 
\frac{\omega^2_n-(\sigma^2-\rho^2) }{(\omega^2_n+\sigma^2-\rho^2 )^2
+(2 \rho \omega_n)^2 } \nonumber \\
&=& \frac{i}{2} \left[ \frac{\sinh(\beta \rho)}{\cosh(\beta \sigma)+
\cosh(\beta \rho )}\right].
\eea
When the GN model is studied in the canonical formalism (i.e., with a fixed
charge)\cite{canonical}, similar calculations appear 
with imaginary chemical potential. In this case a regulating factor 
of the form 
$\e^{i \omega_n \tau}$ is needed in evaluating the summation in \Eq (7) and
ensures a finite result in the limit $\tau \rightarrow 0$ after the Matsubara
sum has been performed. By using \eqs (6) and (7), we obtain 
\be
T \sum^{\infty}_{n=-\infty} \ln[(\omega_n -i\rho )^2 + \sigma^2 ]
=T\left[\ln 2+\ln(\cosh(\beta \sigma)+\cosh(\beta \rho)) \right],
\ee
where the $\zeta$-function regularization was used to determine the 
field-independent constant. At zero temperature, these formulae 
reduce to
\bea
\int \frac{dp_0}{2 \pi }\frac{1}{(p_0 -i \rho )^2+\sigma^2}
&=& \frac{\theta(|\sigma|-|\rho|)}{2|\sigma|}, \\
\int \frac{dp_0}{2 \pi }\frac{p_0-i\rho }{(p_0 -i \rho )^2+\sigma^2}
&=& \frac{i}{2} \sgn(\rho) \theta(|\rho|-|\sigma|), \\
\int \frac{dp_0}{2 \pi }\ln[(p_0 -i \rho )^2+\sigma^2] 
&=& \max (|\sigma|,|\rho|). 
\eea

Using \eqs (6), (7) and $E_\sigma \equiv \sqrt{{\bf p^2}+\sigma^2}$, we have
\bea
\frac{\partial}{\partial \sigma }\left( \frac{V_T}{N}\right)
&=& \frac{\sigma}{g^2}-2\sigma \int \frac{d^{d-1} {\bf p}}{(2\pi )^{d-1}}
\frac{1}{E_\sigma} \left[ 1-\frac{1}{1+\e^{\beta (E_\sigma +\rho )}}
-\frac{1}{1+\e^{\beta (E_\sigma -\rho )}}\right], \\
\frac{\partial}{\partial \rho }\left( \frac{V_T}{N}\right)
&=& \frac{\rho}{R^2}-2\sinh(\beta \rho ) \int \frac{d^{d-1} 
{\bf p}}{(2\pi )^{d-1}}
\frac{1}{\cosh(\beta E_\sigma)+\cosh(\beta \rho)}. 
\eea 

To renormalize the effective potential $V_T (\sigma , \rho)$ let us consider
the GN model at zero temperature and in the absence of the random chemical 
potential because the effects of temperature and chemical potential do not
change the ultraviolet behavior. 
Define $1/G^2 \equiv 1/g^2 - 1/g^2_c $ with $1/g^2_c \equiv 4 \int 
\frac{d^d p}{(2\pi )^d} \frac{1}{p^2 }$. 
In $2<d<4$, the GN model is in the broken phase of the $Z_2$ chiral symmetry 
for negative $G^2$,
corresponding to strong coupling ($g^2 > g^2_c $) in the cutoff regularization,
while it is in the symmetric phase for $G^2 \geq 0$, corresponding to weak
coupling ($ 0<g^2 \leq g^2_c $). In particular, in two dimensions, the $Z_2 $
chiral symmetry of the GN model must be broken no matter how we choose the
coupling $g^2 $\cite{rwprep}. In the broken phase,
\be
\frac{1}{g^2}=4\int \frac{d^d p}{(2 \pi )^d} \frac{1}{p^2 + M^2 },
\ee
where $M= |\langle \sigma \rangle | (>0)$ is the dynamically generated fermion 
mass at zero temperature.
 
From now on, we will adopt dimensional regularization, where $G^2 $ is 
equal to the regularized $g^2 $.

\subsection{Three dimensions}

In this case, renormalization is not
needed to the leading order of the $1/N$ expansion (in dimensional
regularization).
By making use of \Eq (11), we can find the zero-temperature effective
potential $V_0 (\sigma , \rho)$ directly:
\be
\frac{V_0 (\sigma,\rho)}{N} = \frac{\sigma^2 }{2G^2 }+\frac{\rho^2}{2 R^2 }
-\frac{1}{6 \pi} 
\left[ \left.\max\right.^3 (|\sigma|,|\rho|)-3\sigma^2 \max(|\sigma|,|\rho|) \right],
\ee
where $1/G^2 = -M/\pi $ for broken $Z_2$ chiral symmetry. 
The gap equations have four kinds of solution $(|\sigma|,|\rho|)$:
(i) $(0,0)$, (ii) $(M,0)$, (iii) $(0,2\pi /R^2 )$ and
(iv) $(\sqrt{M(M-2\pi /R^2)},M)$. The solution (iv) exists only for
$M > 2\pi/R^2$ and corresponds to saddle points.
Fig. 1 shows the zero-temperature effective potential as a function of 
$\sigma /M$ and $\rho /M$, for broken $Z_2 $ chiral symmetry.
$\langle \rho \rangle =0$ is metastable, irrespective of the values of
$G^2$ and $R^2$. For $|\rho| >|\sigma|$, however,
$V_0 (\sigma, \rho)$ is unbounded from
below due to the $-|\rho|^3 /(6 \pi )$ term, which indicates breaking of
the charge conjugation symmetry. This result stems from the fact that 
the term $-|\rho|^3/(6\pi )$ arising from quantum effects 
surpasses the effect of the 
probability distribution ($ \rho^2 /(2 R^2 )$) for large $|\rho|$. 

At finite temperature, using \eqs (6), (7) and dimensional regularization, we
obtain
\bea
\int^{(T)}_p \frac{1}{(\omega_n - i\rho )^2+E^2_\sigma}
&=& -\frac{1}{4 \pi \beta } \left[\beta |\sigma|+ \ln \left(
1+2\e^{-\beta |\sigma|} \cosh(\beta \rho )+\e^{-2\beta|\sigma|}\right) 
\right], \\
\int^{(T)}_p \frac{\omega_n -i \rho }{(\omega_n - i\rho )^2+E^2_\sigma}
&=& -\frac{i~\sgn (\rho)}{4 \pi \beta^2 }  \left[ 
\beta |\sigma| \ln \left( \frac{1+\e^{\beta (|\sigma|+|\rho|)}}
{1+\e^{\beta (|\sigma|-|\rho|)}}\right) 
+\Li_2(-\e^{\beta (|\sigma|+|\rho|)})
-\Li_2(-\e^{\beta (|\sigma|-|\rho|)}) \right].
\eea 
Here the polylogarithm $\Li_\nu (z)$  
is defined (for $\nu >0$) as $\Li_\nu (z)=\sum^\infty_{k=1} \frac{z^k}{k^\nu}$
(see Ref.\cite{hw} for useful properties).
From these formulae, the finite-temperature effective potential 
$V_T (\sigma,\rho)$ is given by
\bea
\frac{V_T (\sigma,\rho)}{N}&=& \frac{\sigma^2 }{2G^2}+\frac{\rho^2}{2R^2}
-\frac{\sigma^3}{3\pi }
+\frac{1}{\pi \beta^3 }\left[
\Li_3 (-\e^{\beta (\sigma+\rho)})+\Li_3 (-\e^{\beta (\sigma-\rho)})\right.
\nonumber \\
&&\left. -\beta \sigma \left\{ \Li_2 (-\e^{\beta (\sigma+\rho)})
+\Li_2 (-\e^{\beta (\sigma-\rho)})\right\} \right],
\eea
up to a field-independent constant. At sufficiently high temperature,
\be
\frac{V_T (\sigma,\rho)}{N} \approx \left( \frac{\ln 2}{\pi }\right)
T (\sigma^2 -\rho^2 ).
\ee
While the initially broken $Z_2$ chiral symmetry is restored 
at high temperature,
charge conjugation symmetry is not. Hence our model exhibits 
nonrestoration of charge conjugation symmetry irrespective of the values
of $G^2$ and $R^2$. We may interpret
this phenomenon as an inverse symmetry breaking because $\langle \rho \rangle
=0$ is metastable at zero temperature. Intuitively, SNR is related to 
the tachyon-like behavior of the random chemical potential
(see \eqs (19) and (24)). In the quantum
correction term of \Eq (5) the chemical potential acts 
as a negative mass term ($-\rho^2$) 
contrary to the usual positive mass term ($\sigma^2$).
From a different point of view, we will discuss the origin of SNR conceptually
in Sec. IV.

\subsection{Two dimensions}

For dimensional regularization, we work in $2+\epsilon$ dimensions. 
In terms of the fermion
mass $M$, the zero-temperature effective potential is given by
\bea
\frac{V_0 (\sigma,\rho)}{N}&=&\frac{\sigma^2}{2\pi }\left[-1 +
2 \ln \left( \frac{\max(|\sigma|,|\rho|)
+\sqrt{\max^2 (|\sigma|,|\rho|)-\sigma^2}}{M}\right)  \right] \nonumber \\
&&+\frac{\rho^2}{2R^2}-\frac{|\rho|}{\pi }
\sqrt{\left.\max\right.^2 (|\sigma|,|\rho|)-\sigma^2},
\eea
where \eqs (11) and (14) were used. Unlike in three dimensions, for large 
$|\rho|$ the effect of the probability distribution ($\rho^2 /(2R^2 )$)
is comparable to the last term ($\approx -\rho^2 /\pi $) in \Eq (20) 
arising from quantum effects.
The charge conjugation symmetry can be
controlled by the strength of disorder $R^2$. 
The system is in the symmetric state
for $0<R^2 < \pi /2$, while in the broken state for $R^2 > \pi /2$.
Fermions and antifermions are equally probable in the symmetric state 
($\langle \rho \rangle =0$), but only fermions (or antifermions) are allowed 
in the broken state ($\langle \rho \rangle = \pm \infty$).
Our system suffers from a quantum phase transition 
at $R^2 =\pi /2(\equiv R^2_c )$.
The gap equations have solutions $(|\sigma |,|\rho |)$: 
(i) $(0,0)$, (ii) $(M,0)$ and (iii) $(\sqrt{(2R^2-\pi )/(2R^2 +\pi )} M,
2R^2 M/(2R^2 +\pi ))$ for $R^2 \neq R^2_c$, and (i) $(0, \forall |\rho |)$ and
(ii) $(M,0)$ for $R^2 =R^2_c$. 
The solution (iii) exists only for $R^2 >R^2_c$ and
corresponds to saddle points.
Fig. 2 shows the zero-temperature effective potential as a function of 
$\sigma/M$ and $\rho/M$ in (a) the symmetric and 
(b) the broken phase for the charge
conjugation symmetry.

To examine the high-temperature ($\beta \rightarrow 0$) behavior, let us 
introduce dimensionless quantities: $\tilde{V}_T=\beta^2 V_T ,~\ts
=\beta \sigma,~\tr =\beta\rho,~\tilde{M}=\beta M$. We want to expand
the finite-temperature effective potential in $\ts $ and 
$\tr $ at high temperature (i.e., for small $\ts $ and 
$\tr $). In terms of the dimensionless quantities, \eqs (12) and (13)
are reduced to
\bea
\frac{\partial}{\partial \ts } \left( \frac{\tilde{V}_T}{N} \right)
&=&\frac{2 \ts }{\pi } \left[ \ln \left( \frac{|\ts | }{\tilde{M} }\right)
+\int^\infty_0 dx \frac{1}{\sqrt{x^2+\ts^2 }}\left( 
\frac{1}{1+\e^{\sqrt{x^2+\ts^2}+\tr }} +
\frac{1}{1+\e^{\sqrt{x^2+\ts^2}-\tr }} \right)\right] \nonumber \\
&=&\frac{2\ts }{\pi }\left[ \left\{ \ln \left( \frac{\pi }{\tilde{M} }\right)
-\gamma+{\rm O}(\ts^2) \right\}
+\left\{ \frac{7\zeta(3)}{4\pi^2 } +{\rm O}(\ts^2) \right\} \tr^2 
+{\rm O}(\tr^4) \right], \\
\frac{\partial}{\partial \tr } \left( \frac{\tilde{V}_T}{N} \right)
&=& \tr \left[ \frac{1}{R^2}-\frac{2\sinh(\tr)}{\pi \tr } 
\int^\infty_0 dx \frac{1}{\cosh(\sqrt{x^2+\ts^2})+\cosh(\tr)} 
\right]\nonumber \\
&=&\tr \left[ \frac{1}{R^2 }-\frac{2}{\pi }+\left\{ \frac{7\zeta(3)}{2\pi^3 }
+{\rm O}(\tr^2)\right\} \ts^2
+{\rm O}(\ts^4) \right].
\eea
To obtain the O($\tr^0$) term in the bracket of \Eq (21) we used the 
integration formula\cite{dj} for small $\ts^2$: 
\be
\int^\infty_0 dx \frac{1}{\sqrt{x^2+\ts^2}\left( 1+\e^{\sqrt{x^2+\ts^2}}
\right)} = -\frac{1}{2} \left[ \ln \left(\frac{|\ts|}{\pi}\right)+\gamma
+{\rm O}(\ts^2)\right].
\ee
Integrating \eqs (21) and (22), at sufficiently high temperature, we obtain
\be
\frac{V_T (\sigma,\rho)}{N}=\frac{1}{\pi }\left[ \ln \left( 
\frac{\pi T}{M}\right)-\gamma \right]\sigma^2 +\frac{1}{2} \left( 
\frac{1}{R^2}-\frac{2}{\pi} \right)\rho^2 +\frac{7 \zeta(3)}{4 \pi^3 T^2} 
\sigma^2 \rho^2 + \cdots ,
\ee
up to a field-independent constant. Since the O($\ts^0$) term in the bracket
of \Eq (22) is exact, $V_T (\sigma, \rho)$ has no term higher than the
$\rho^2$ term that consists of $\rho$ fields only. At high temperature,
the $Z_2$ chiral symmetry is always restored. However, the behavior of the
charge conjugation symmetry is the same as that at zero temperature. Hence,
charge conjugation symmetry is not restored at high temperature,
for $R^2 >R^2_c$.

Until now in order to find the effective potential $V_T (\sigma,\rho)$
we have used the mean field approximation (MFA) by taking $\sigma$ and $\rho$
as constant fields. This corresponds to the leading approximation in the
$1/N$ expansion where the $\sigma$ and $\rho$ loops (i.e., the fluctuations
of the $\sigma$ and $\rho$ fields) are not included. 
Let us check the validity of our calculations.
In the large $N$ limit MFA is good, while for large but finite $N$ 
it may fail due to the contribution from kinks in two dimensions\cite{dmr,kkw}.

At first consider the case of the usual two-dimensional GN model 
(without the random chemical potential) where MFA predicts a wrong critical
temperature $T_0 (\neq 0)$\cite{dmr}. 
For $0 \leq T <T_0$ the MFA effective potential
is the double well with two degenerate minima at $M(T)$ and $-M(T)$, the 
solutions of the gap equation. Due to quantum tunneling between two
degenerate minima, the system has kink solutions alternating between $M(T)$
and $-M(T)$. They have higher energies than the 
constant solutions $M(T)$ and $-M(T)$. The Helmholtz free energy $F$ is 
related to the internal energy $U$ and the entropy $S$ by $F=U-TS$.
Since $F=U$ at zero temperature, the constant solution is favored and MFA
is expected to be valid. If we consider the contribution from kinks
explicitly, we find the average (or blocking) potential\cite{ap} which has
a plateau between $M(\equiv M(0))$ and $-M$ \cite{dmr}. 
Even in the presence of an
arbitrarily small external field this potential is tilted to favor $M$ or
$-M$. Hence the contribution from kinks does not change the physics materially
and MFA is qualitatively valid at zero temperature. For $0<T<T_0$ the number
of kink configurations is sufficiently large to gain enough entropy and so
their probability is overwhelming. Since the region of $\sigma=M(T)$ will,
on the average, have the same weight as those of $\sigma=-M(T)$, we have 
$\langle \sigma \rangle =0$, which indicates the breakdown of MFA. 
For $T\geq T_0$ the system is in the symmetric phase ($M(T)=0$) in MFA and
thus has no kink solutions. Consequently, for the two-dimensional GN model
MFA is good only at zero and high temperature ($T\geq T_0$) and, by the 
formation of kinks, the true critical temperature turns out to be zero.

Now let us examine the validity of MFA for the two-dimensional GN model with
the random chemical potential \eqs (3) or (4). For the purpose of the present
paper we consider only the cases of zero and sufficiently high
temperature. At zero temperature the MFA effective potential 
$V_0 (\sigma,\rho)$ has degenerate minima at ($|\sigma|,|\rho|$): ($M,0$)
for $0<R^2 \leq R^2_c$ and ($0,\infty$) for $R^2 >R^2_c$ (see Fig. 2 (a) and
(b)). Hence for $0<R^2\leq R^2_c$ our system may have kink solutions for
$\sigma$ alternating between $M$ and $-M$, but no kinks for $\rho$. 
In this case the situation is similar to that of 
the usual GN model in the previous paragraph. Thus it is expected that MFA is
good at zero temperature and for $0<R^2\leq R^2_c$. For $R^2 >R^2_c$ the MFA
effective potential is unbounded from below and there is no tunneling 
between two degenerate ground states because of infinitely high barrier.
So $\sigma$ and $\rho$ do not have kink solutions. At sufficiently high
temperature the $\sigma$ and $\rho$ fields are decoupled from each other in  
$V_T (\sigma,\rho)$ and can be treated separately. By the restoration of the
$Z_2$ chiral symmetry in MFA, $\sigma$ has no
kink solutions, irrespective of the value of the disorder strength. 
For $0<R^2<R^2_c$ the charge conjugation symmetry is preserved 
in MFA and so no kinks for $\rho$. 
For $R^2>R^2_c$ the situation is the same as that at zero temperature. 
For $R^2=R^2_c$ the MFA effective potential for $\rho$ vanishes.
As $R^2$ is tuned through $R^2_c$, the charge conjugation symmetry undergoes
a first-order phase transition from the symmetric phase 
($\langle \rho \rangle =0$) to the broken phase ($\langle \rho \rangle 
=\infty$ or $-\infty$), 
following positive or negative values of $\langle \rho \rangle$ 
according to the value of $\langle \rho \rangle$ in the broken
phase. So it is reasonable to assume that $\rho$ has no
kink solution at $R^2=R^2_c$. As a result, MFA is reliable at zero 
and high temperature for all values of the disorder strength.

\section{Origin of charge conjugation symmetry nonrestoration}

In this section we discuss the mechanism of SNR for the charge conjugation
symmetry conceptually. For convenience 
set $\sigma=0$ in \eqs (4) and (5) because SNR 
is the effect of the random chemical potential. That is, we neglect the 
four-fermion interaction $(\bar{\Psi}\Psi)^2$ and consider the system of
free massless Dirac fermions in the presence of the random chemical 
potential. To leading
order in the $1/N$ expansion the finite-temperature effective potential 
$V_T (0,\rho)/N$ consists of two parts: (i) a term from probability
distribution ($\rho^2 /(2R^2)\equiv P_R (\rho)$) 
and (ii) the grand free energy 
(or grand potential) for free massless Dirac fermions at constant 
chemical potential $\rho$ ($\equiv \Omega_T (\rho)$).

So it is essential to conceptually determine the value of the chemical 
potential that $\Omega_T (\rho)$ favors. 
By the symmetry of $\Omega_T (\rho)$ we can
restrict ourselves to the positive chemical potential without loss of
generality. At first, consider the case of zero temperature. According to
Fermi-Dirac statistics, fermions fill all the energy levels to the Fermi
energy (=chemical potential) and antifermions are suppressed. Hence, the larger
chemical potential, the larger charge (number) density
(= fermion number density$-$antifermion number density). This result is 
retained at nonzero temperature. Since the grand free energy density is minus
the pressure\cite{LL}, it is a decreasing function of the (positive) 
charge density. Consequently, the large charge 
density (i.e., large chemical potential) is preferred 
and, for all temperatures, 
$\Omega_T (\rho)$ is minimized at large chemical potential
($|\rho| \rightarrow \infty$). 
This implies SNR for the charge conjugation symmetry 
in the open system of free massless Dirac fermions. Moreover, we can guess the
functional form of $\Omega_T (\rho)$ by dimensional analysis: At zero
temperature, $\Omega_0 (\rho) \propto -|\rho|^d$ in $d$ dimensions. 
At sufficiently high temperature, $\Omega_T (\rho) \propto -T\rho^2 
+{\rm O}(\rho^4/T)$ in three dimensions and $\Omega_T (\rho) \propto -\rho^2
+{\rm O}(\rho^4 /T^2)$ in two dimensions, up to a field-independent constant.
These qualitative results can be checked 
explicitly from \eqs (15), (19), (20) and (24). 

Now let us consider the contribution $P_R (\rho)$ 
from the probability distribution of the
random chemical potential. In three dimensions, for large $|\rho|$, 
$\Omega_T (\rho)$ exceeds $P_R (\rho)$ at zero and sufficiently high 
temperature, irrespective of the value
of the disorder strength. Therefore, the initially
broken charge conjugation symmetry is not restored at high temperature. In two
dimensions $P_R (\rho)$ is comparable to $\Omega_T (\rho)$. The probability 
distribution of the random chemical potential for weak disorder (small $R^2$) 
is dominated at $\rho=0$ and the charge conjugation symmetry is preserved at
zero and high temperature. 
However, for strong disorder (large $R^2$) all values of the chemical 
potential have small probability density and so the fate of the 
charge conjugation symmetry is determined by $\Omega_T (\rho)$. 
Thus, in this case, the initially broken charge conjugation symmetry 
is not restored at high temperature.    

\section{Conclusions}

In the present paper we examined the symmetry behavior of 
the Gross-Neveu model with
random chemical potential which is equivalent to the four-fermion model 
\Eq (3). We used the leading approximation in the $1/N$ expansion 
(i.e., the mean field approximation).
Our model has the charge conjugation symmetry as well as the $Z_2$ chiral
symmetry. The initially broken $Z_2$ chiral symmetry is always restored 
at high temperature.
In three dimensions, the charge conjugation symmetry that is broken at zero
temperature, is not restored at high temperature, 
irrespective of the value of the disorder strength $R^2$. 
In two dimensions, at zero temperature, the charge conjugation
symmetry is not broken for weak disorder ($0<R^2<R^2_c (=\pi/2)$), but broken 
for strong disorder ($R^2 >R^2_c$). 
Therefore, our system exhibits a quantum phase transition at $R^2 =R^2_c$ 
as the value of $R^2$ is varied. 
For any given value of $R^2$ the high-temperature behavior 
of the charge conjugation symmetry is the same as its zero-temperature 
behavior. Hence the charge conjugation symmetry remains broken at high 
temperature (i.e., symmetry nonrestoration) for $R^2 >R^2_c$.
By examining the existence of the kink solutions we checked that the mean
field approximation is reliable even in two dimensions at zero and high
temperature.

In addition, we discussed our results on charge conjugation symmetry 
nonrestoration conceptually, after neglecting the four-fermion interaction
$(\bar{\Psi} \Psi)^2$ for convenience because symmetry nonrestoration is the
effect of the random chemical potential. The behavior of the charge
conjugation symmetry is determined by the competition of two terms
in the finite-temperature effective potential: (i) the term $\rho^2/(2R^2)$
from the Gaussian distribution for the chemical potential $\rho$ that favors
the symmetric phase ($\langle \rho \rangle =0$) 
and (ii) the grand free energy for free
massless Dirac fermions which favors the broken phase 
($|\langle \rho \rangle| \rightarrow \infty$).

As further work, it would be worthwhile to perform next-to-leading order 
calculations
and consider a non-Gaussian distribution for the random chemical potential.

\section*{ACKNOWLEDGMENTS}

We thank Dr. P. Vranas for helpful conversations and careful reading of the
manuscript. This work was supported
in part by the National Science Foundation under grant NSF-PHY96-05199.
S.I.H. was also supported by the Korea Science and Engineering Foundation.

\newpage
\begin{figure}
\caption{The zero-temperature effective potential $V_0 /(NM^3)$ in three
dimensions as a function of $\sigma/M$ and $\rho/M$ for $R^2 M =10$, in the
case of the $Z_2 $ chiral symmetry breaking ($G^2 M=-\pi $).}
\end{figure}
\begin{figure}
\caption{The zero-temperature effective potential $V_0 /(NM^2)$ in two
dimensions as a function of $\sigma/M$ and $\rho/M$, 
for (a) $R^2 =1$ and (b) $R^2 =10$.}
\end{figure}

\end{document}